\documentclass{svmult}

\usepackage{makeidx}         
\usepackage{graphicx}        
\usepackage{multicol}        
\usepackage[bottom]{footmisc}

\makeindex             

\begin{document}

\title*{Uncovering the Internal Structure of the Indian Financial Market:\\
{\Large Cross-correlation behavior in the NSE}}
\titlerunning{Internal Structure of the Indian Market} 
\author{Sitabhra Sinha and Raj Kumar Pan}
\institute{The Institute of Mathematical Sciences, C. I. T. Campus,
Taramani,\\ 
Chennai - 600 113, India.\\
\texttt{sitabhra@imsc.res.in}}

\maketitle
The cross-correlations between price fluctuations of 201 frequently traded
stocks in the National Stock Exchange (NSE) of India are analyzed in this
paper. We use daily closing prices for the period 1996-2006, which
coincides with the period of rapid transformation of the market following
liberalization. The eigenvalue distribution of the cross-correlation
matrix, $\mathbf{C}$, of NSE is found to be similar to that of developed
markets, such as the New York Stock Exchange (NYSE): the majority of
eigenvalues fall within the bounds expected for a random matrix constructed
from mutually uncorrelated time series. Of the few largest eigenvalues that
deviate from the bulk, the largest is identified with market-wide movements.
The intermediate eigenvalues that occur between the largest
and the bulk have been associated in NYSE with specific business
sectors with strong intra-group interactions.
However, in the Indian market, these deviating eigenvalues are comparatively 
very few
and lie much closer to the bulk. We propose that this is because of the
relative lack of distinct sector identity in the market, with the movement 
of stocks
dominantly influenced by the overall market trend. This is shown
by explicit construction of the interaction network in the market,
first by generating the minimum spanning tree from the unfiltered correlation
matrix, and later, using an improved method of generating the graph
after filtering out the market mode and random effects from the data.
Both methods show, compared to developed markets, the relative absence 
of clusters of co-moving stocks that belong to the same business sector.
This is consistent
with the general belief that emerging markets tend to be more correlated
than developed markets. 
\section{Introduction}
\label{sec:1}
\begin{quote}
``Because nothing is completely certain but subject to fluctuations, 
it is dangerous for people to allocate their capital to a single or 
a small number of securities. [\ldots] No one has reason to expect that all
securities \ldots will cease to pay off at the same time, 
and the entire capital be lost.'' -- {\em from the 1776 prospectus of an
early mutual fund in the Netherlands}~\cite{rouwenhorst05}
\end{quote}
As evident from the above quotation, the correlation between price
movements of different stocks has long been a topic of vital interest to
those involved with the study of financial markets. With the recent
understanding of such markets as examples of complex systems with many
interacting components, these cross-correlations have been used to infer
the existence of collective modes in the underlying dynamics of stock
prices.  It is natural to expect that stocks which strongly interact with
each other will have correlated price movements. Such interactions may
arise because the companies belong to the same business sector (i.e., they
compete for the same set of customers and face similar market conditions),
or they may belong to related sectors (e.g., automobile and energy sector
stocks would be affected similarly by rise in gasoline prices), or they may
be owned by the same business house and therefore perceived by investors to
be linked. In addition, all stocks may respond similarly to news breaks
that affect the entire market (e.g., the outbreak of a war) and this
induces market-wide correlations. On the other hand, information that is
related only to a particular company will tend to decorrelate its price
movement from those of others.

Thus, the effects governing the cross-correlation behavior of stock price
fluctuations can be classified into (i) market (i.e., common to all
stocks), (ii) sector (i.e., related to a particular business sector) and
(iii) idiosyncratic (i.e., limited to an individual stock). The empirically
obtained correlation structure can then be analyzed to find out the
relative importance of such effects in actual markets.  Physicists
investigating financial market structure have focussed on the spectral
properties of the correlation matrix, with pioneering studies investigating
the deviation of these properties from those of a random matrix, which
would have been obtained had the price movements been uncorrelated. It was
found that the bulk of the empirical eigenvalue distribution matches fairly
well with those expected from a random matrix, as does the distribution of
eigenvalue spacings~\cite{laloux99,plerou99}. Among the few large
eigenvalues that deviated from the random matrix predictions, the largest
represent the influence of the entire market common to all stocks, while
the remaining eigenvalues correspond to different business
sectors~\cite{gopikrishnan01}, as indicated by the composition of the
corresponding eigenvectors~\cite{plerou02}. However, although models in
which the market is assumed to be composed of several correlated groups of
stocks is found to reproduce many spectral features of the empirical
correlation matrix~\cite{noh00}, one needs to filter out the effects of the
market-wide signal as well as noise in order to identify the group
structure in an actual market. Recently, such filtered matrices have been
used to reveal significant clustering among a large number of stocks from
the NYSE~\cite{kim05}.

The discovery of complex market structure in developed financial
markets as NYSE and Japan~\cite{utsugi04}, brings us to the question of
whether emerging markets show similar behavior. While it is generally
believed that stock prices in developing markets tend to be relatively more
correlated than the developed ones~\cite{morck00}, there have been very few
studies of the former in terms of analysing the spectral properties of
correlation matrices~\cite{wilcox04,kulkarni05,jung06,cukur07,sinha06} 
\footnote{Most studies of
correlated price movements in emerging markets have looked at {\em
synchronicity} which measures the incidence of similar (i.e., up or down)
price movements across stocks, and is not the same as correlation which
measures relative magnitude of the change as well as its direction,
although the two are obviously closely related.}.

In this paper we present the first detailed study of cross-correlations in
the Indian financial market over a significant period of time, that
coincides with the decade of rapid transformation of the recently
liberalized economy into one of the fastest growing in the world.  The
prime motivation for our study of one of the largest emerging markets is to
see if there are significant deviations from developed markets in terms of
the properties of its collective modes. As already shown by
us~\cite{sinha06,pan06,pan06a} the return distribution in Indian markets
follows closely the ``inverse cubic law" that has been reported in
developed markets. If therefore, deviations are observed in the
correlation properties, these would be almost entirely due to differences
in the nature of interactions between stocks. Indeed, we do observe
that the Indian market shows a higher degree of correlation compared to,
e.g., NYSE. We present the hypothesis that this is due to the
dominance of the market-wide signal and relative absence of significant
group structure among the stocks. This may indicate that one of the
hallmarks of the transition of a market from emerging to developed status
is the appearance and consolidation of distinct business sector identities.

\section{The Indian Financial Market}
\label{sec:2}
There are 23 different stock markets in India. The largest of these is the
National Stock Exchange (NSE) which accounted for more than half of the
entire combined turnover for all Indian financial markets in
2003-04~\cite{ismr}, although its market capitalization is comparable to
that of the second largest market, the Bombay Stock Exchange. The NSE is
considerably younger than most other Indian markets, having commenced
operations in the capital (equities) market from Nov 1994.  However, as of
2004, it is already the world's third largest stock exchange (after NASDAQ
and NYSE) in terms of transactions~\cite{ismr}.  It is thus an excellent
source of data for studying the correlation structure of price movements in
an emerging market.

{\em Description of the data set.} We have considered the daily closing
price time series of stocks traded in the NSE available from the exchange
web-site~\cite{website}.  For cross-correlation analysis, we have focused
on daily closing price data of $N =201$ NSE stocks from Jan 1, 1996 to May
31, 2006, which corresponds to $T = 2607$ working days (the individual
stocks, along with the business sector to which they belong, are given
in Table~\ref{ss:table1}).  The selection of
the stocks was guided by the need to minimise missing data in the
time-series, a problem common to data from other emerging
markets~\cite{wilcox04}. In our data, 45 stocks have no missing data, while
from the remaining stocks, the one having the largest fraction of missing
data has price data missing for less than $6 \%$ of the total period
covered \footnote{In case of a date with missing price data, it is assumed
that no trading took place on that day, so that, the price
remained the same as the preceding day.}.

\section{The Return Cross-Correlation Matrix}
\label{sec:4}
To measure correlation between the price movements across different
stocks, we first need to measure the price fluctuations such that the 
result is independent of the scale of  measurement. For this, we 
calculate the logarithmic return of price. 
If $P_{i}(t)$ is the stock price of the $i$th stock at time $t$, then the
(logarithmic) price return is defined as
\begin{equation}
R_{i}(t,\Delta t) \equiv \ln {P_{i}(t+\Delta t)}- \ln {P_{i}(t)}. \label{return}
\end{equation}
For daily return, $\Delta t$ = 1 day.
By subtracting the average return and dividing the result with the
standard deviation of the returns (which is a measure of the volatility
of the stock),
$ \sigma_{i} = \sqrt{\langle R_{i}^{2} \rangle - \langle R_{i} \rangle^{2}}$,
we obtain the normalized price return,
\begin{equation}
r_{i}(t,\Delta t) \equiv \frac{R_{i}-\langle R_{i} \rangle}{\sigma_{i}},
\label{normalized_return}
\end{equation}
where $\langle \ldots \rangle$ represents time average. Once the
return time series for $N$ stocks over a period of $T$ days are obtained,
the cross-correlation matrix ${\mathbf C}$ is calculated, whose element 
$C_{ij}=\langle r_{i} r_{j} \rangle$,
represents the correlation between returns for stocks $i$ and $j$.

\begin{figure}[tbp]
\centering
\includegraphics[width=0.49\linewidth,clip]{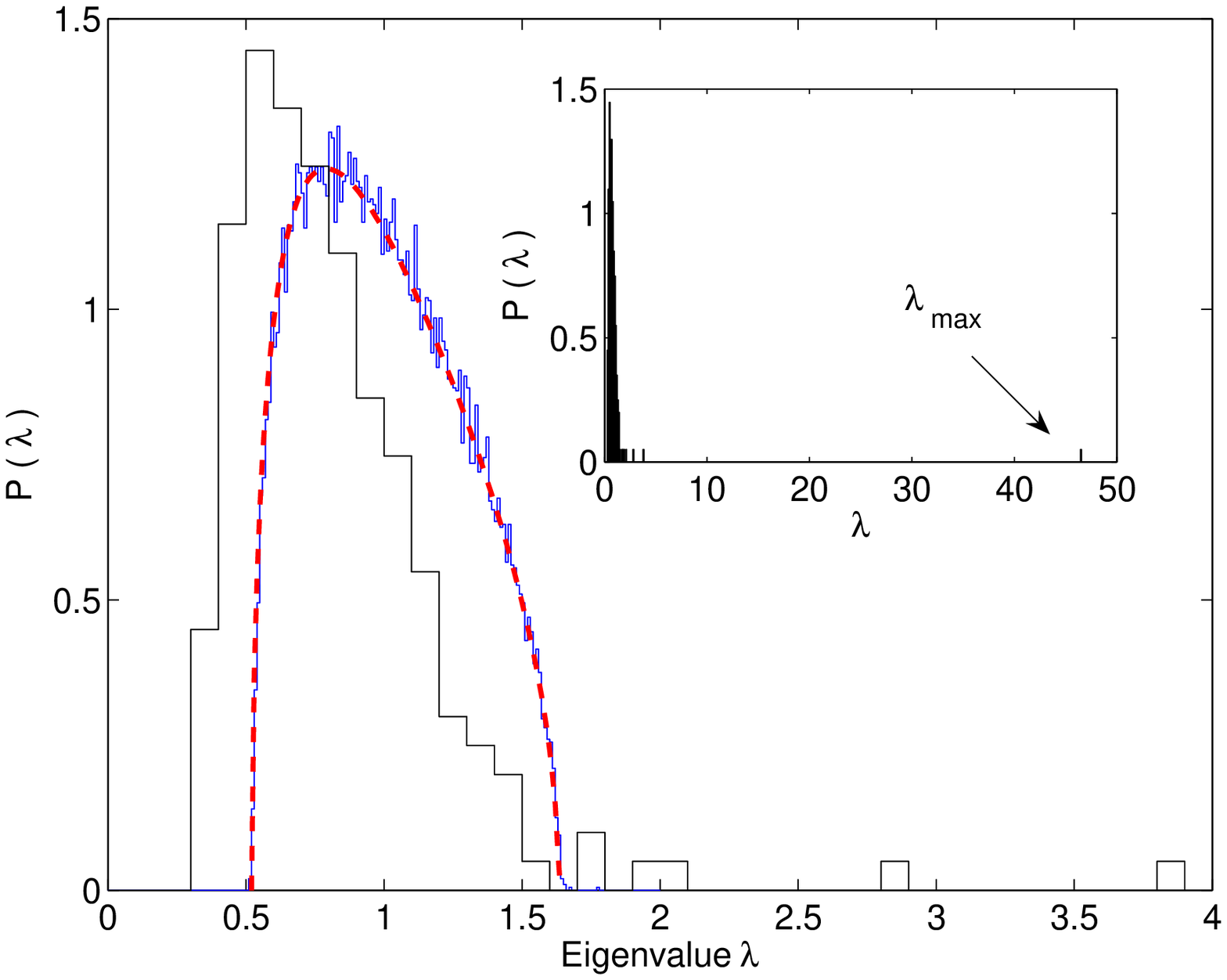}
\includegraphics[width=0.49\linewidth,clip]{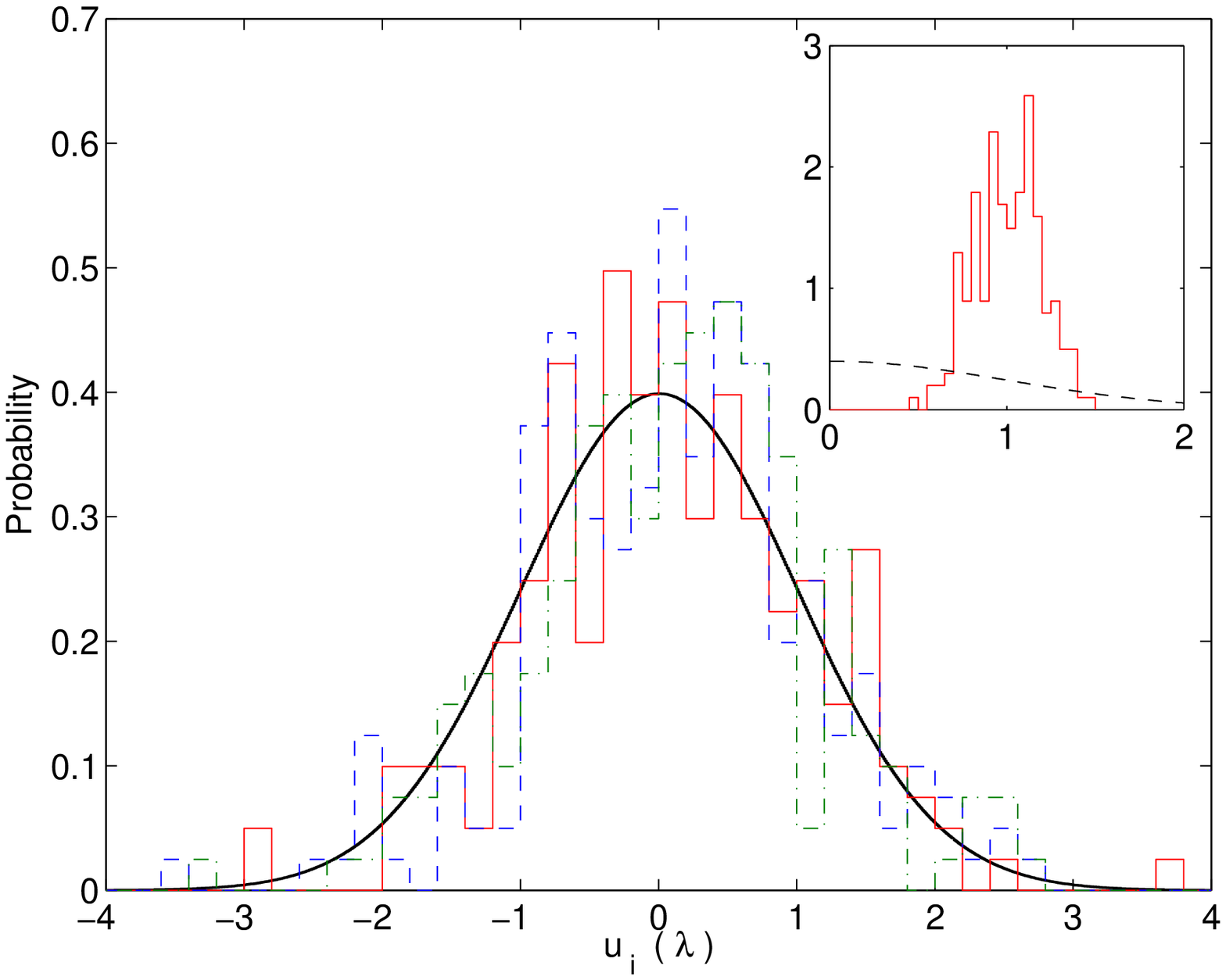}\\
\caption{(left) The probability density function of the eigenvalues of the 
cross-correlation matrix ${\mathbf C}$ for 201 stocks in the NSE of India
for the period Jan 1996-May 2006. For comparison
the theoretical distribution predicted by Eq.~(\ref{eq:sengupta}) is shown
using broken curves, which overlaps with the spectral distribution of the
surrogate correlation matrix generated by randomly shuffling the time series.
The inset shows the largest eigenvalue corresponding to the market.
(Right) The distribution of eigenvector components corresponding to three
eigenvalues belonging to the bulk predicted by RMT and (inset) corresponding
to the largest eigenvalue. In both cases, the Gaussian distribution expected
from RMT is shown for comparison.}
\label{ss:rmt}       
\end{figure}

If the time series are uncorrelated, then the resulting random correlation
matrix, also known as a Wishart matrix, has eigenvalues distributed 
according to~\cite{sengupta99}:
\begin{equation}
P ( \lambda ) = \frac{Q}{2 \pi} 
\frac{\sqrt{(\lambda_{max} - \lambda)(\lambda - \lambda_{min})}}{\lambda},
\label{eq:sengupta}
\end{equation}
with $N \rightarrow \infty$, $T \rightarrow \infty $ such that 
$Q = T / N \geq 1$.
The bounds of the distribution are given by 
$\lambda_{max} = [1 + (1/\sqrt{Q})]^2$ and $\lambda_{min} = 
[1 - (1/\sqrt{Q})]^2$. For the NSE data, $Q = 12.97$, which implies that
the distribution should be bounded at $\lambda_{max} = 1.63$ in the
absence of any correlations. As seen
in Fig.~\ref{ss:rmt}~(left), the bulk of the empirical eigenvalue distribution
indeed occurs below this value.
However, a small fraction ($\simeq 3~\%$) of the eigenvalues
deviate from the random matrix behavior, and, by analyzing them we should
be able to obtain an understanding of the interaction structure of the market.

The random nature of the smaller eigenvalues is also indicated by
an observation of the distribution of the corresponding eigenvector
components. Note that, these components
are normalized for each eigenvalue $\lambda_{j}$ such that,
$\sum_{i=1}^{N}[u_{ji}]^2=N$, where $u_{ji}$ is the $i$-th
component of the $j$th eigenvector. For random matrices generated from
uncorrelated time series, the 
distribution of the eigenvector components is given by the
Porter-Thomas distribution,
\begin{equation} 
P(u)=\frac{1}{\sqrt{2\pi}}\exp-\frac{u^2}{2}.
\label{ss:pt}
\end{equation}
As shown in Fig.~\ref{ss:rmt}~(right), this distribution fits 
the empirical histogram of the eigenvector components for the eigenvalues
belonging to the bulk. However, the eigenvectors of the largest eigenvalues
(e.g., the largest eigenvalue $\lambda_{max}$, as shown in the inset) deviate
quite significantly, indicating their non-random nature.

\begin{figure}[tbp]
\centering
\includegraphics[width=0.95\linewidth,clip]{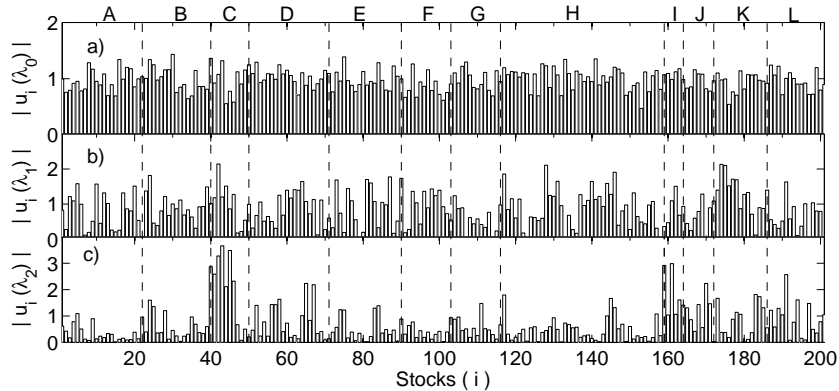}
\caption{The absolute values of the eigenvector components $u_i (\lambda)$
for the three largest eigenvalues of the correlation matrix ${\mathbf C}$.
The stocks $i$ are arranged by business sectors separated by broken lines. 
A: Automobile \& transport, B: Financial, C: Technology, D: Energy, 
E: Basic materials, F: Consumer goods, G: Consumer discretionary,
H: Industrial, I: IT \& Telecom, J: Services, 
K: Healthcare \& Pharmaceutical, L: Miscellaneous.}
\label{ss:eigenvector}       
\end{figure}

The largest eigenvalue $\lambda_0$ for the NSE cross-correlation matrix
is more than 28 times 
larger than the maximum predicted by random matrix theory (RMT).
The corresponding eigenvector
shows a relatively uniform composition, with all stocks contributing to it
and all elements having the same sign (Fig.~\ref{ss:eigenvector}, top).
As this is indicative of a common component that affects all the 
stocks with the same bias, the largest eigenvalue is associated with 
the market mode, i.e., the collective response of the entire market
to information (e.g., newsbreaks)~\cite{laloux99,plerou99}. 

Of more interest for understanding the market structure are the intermediate
eigenvalues that occur between the largest eigenvalue and the bulk 
predicted by RMT. For the NYSE,
it was shown that corresponding eigenvectors of these eigenvalues
are localized, i.e., only a small number of stocks contribute 
significantly to these modes~\cite{gopikrishnan01,plerou02}. It was also 
observed that, for a particular eigenvector, the significantly contributing
elements were stocks that belonged to similar or related businesses
(with the exception of the second largest eigenvalue, where the 
contribution was from stocks having large market capitalization).
Fig.~\ref{ss:eigenvector} shows the stocks, arranged into groups according
to their business sector, contributing to the different intermediate 
eigenvectors
very unequally\footnote{The significant contributions to the second largest
eigenvalue were found to be from the stocks SBIN, SATYAMCOMP, SURYAROSNI, 
ITC, BHEL, NAGARFERT, ACC, GLAXO, DRREDDY and RANBAXY.}. 
For example, it  is apparent that Technology stocks
contribute significantly to the eigenvector corresponding to the third
largest eigenvalue.
However, direct inspection of eigenvector composition for the deviating 
eigenvalues 
does not yield a straightforward interpretation of the significant group
of stocks, possibly because the largest eigenmode
corresponding to the market dominates over all intra-group correlations.

\begin{figure}[tbp] \centering
\includegraphics[width=0.7\linewidth,clip]{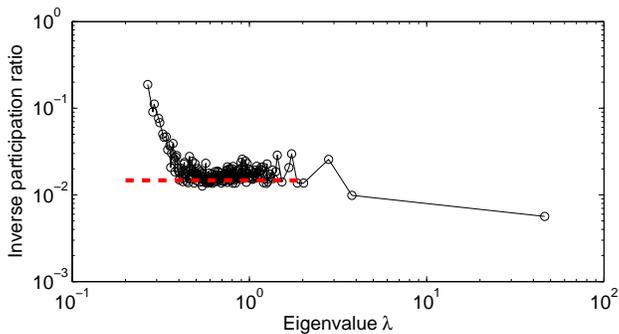}
\caption{Inverse participation ratio (IPR) for the different eigenvalues
of the NSE cross-correlation matrix. The broken line
showing IPR = $3/N$ ($N$ = 201, is the number of stocks) is the expected value 
for a random matrix constructed from $N$ mutually uncorrelated time series.}
\label{ss:ipr}       
\end{figure}
For more detailed analysis of the eigenvector composition, we use the
inverse participation ratio (IPR), which is defined for the $j$-th eigenvector as 
$I_j = \sum_{i=1}^N [ u_{ji}]^4$, where $u_{ji}$ are the component of
$j$th eigenvector. For an eigenvector with equal components, 
$u_{ji} = 1/\sqrt{N}$,
which is approximately the case for the eigenvector corresponding to the
largest eigenvalue, $I_{j} = 1/N$. If, on the other hand, a single component
has a dominant contribution, e.g., $u_{j1} = 1$ and $u_{ji} = 0$ for 
$i \neq 1$, 
we have $I_{j} = 1$. Therefore, IPR is inversely related to the number of
significantly contributing eigenvector components. For the eigenvectors
corresponding to eigenvalues of a random correlation matrix, $\langle I
\rangle \simeq 3/N$.
As seen from Fig.~\ref{ss:ipr}, the eigenvalues belonging
to the bulk predicted by random matrix theory indeed have eigenvectors
with this value of IPR. But, at the lower
and higher end of eigenvalues, the
market shows deviations from this value, suggesting the existence
of localized eigenvectors\footnote{The deviations for the smallest
eigenvalues indicate strong correlations between a few stocks (see
Table~\ref{ss:table2}).}. These deviations are, however, much less significant
and far fewer in number in the Indian market compared to developed
markets, implying that
while correlated groups of stocks do exist in the latter, their existence is
far less clear in the NSE. 

\begin{figure}[tbp] \centering
\includegraphics[width=0.8\linewidth,clip]{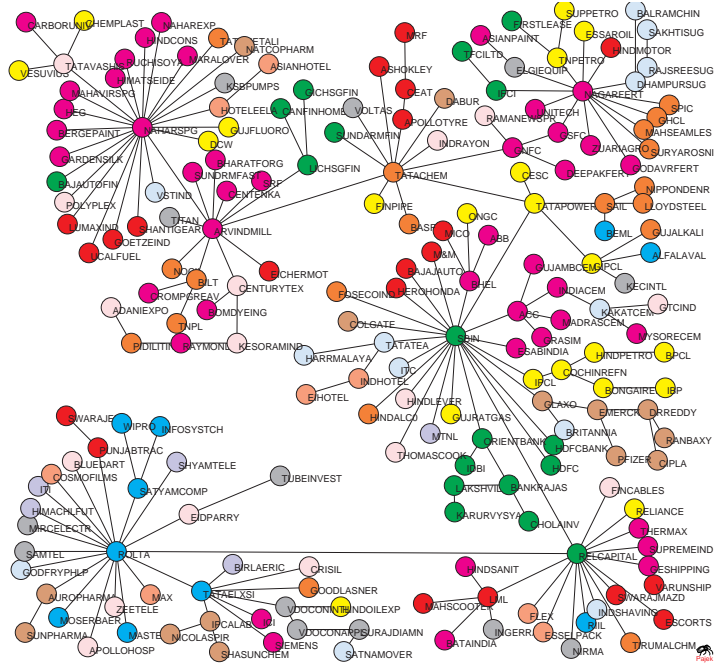}
\caption{The minimum spanning tree connecting 201 stocks of NSE. The node
colors indicate the business sector to which a stock belongs.
The figure has been drawn using the Pajek software.}
\label{ss:mst}       
\end{figure}

In order to graphically present the interaction structure of the stocks in
NSE, we use a method suggested by Mantegna~\cite{Mantegna99} to transform
the correlation between stocks into distances to produce a connected
network in which co-moving stocks are clustered together.
The distance $d_{ij}$ between two stocks $i$ and $j$ are calculated from the 
cross-correlation matrix ${\mathbf C}$, according to 
$d_{ij} = \sqrt{2 (1 - C_{ij})}$. These are used to construct a minimum
spanning tree, which connects
all the $N$ nodes of a network with $N-1$ edges such that the
total sum of the distance between every pair of nodes, $\sum_{i,j} d_{ij}$, is
minimum. For the NYSE, such a construction has been shown to 
cluster together stocks belonging to the same business sector~\cite{Onnela02}.
However, as seen in Fig.~\ref{ss:mst}, for the NSE, such a method fails
to clearly segregate any of the business sectors. Instead, stocks
belonging to very different sectors are equally likely to be found
within each cluster. This suggests that the market mode
is dominating over all intra-sector interactions.

\begin{figure}[tbp]
\centering
\includegraphics[width=0.49\linewidth,clip]{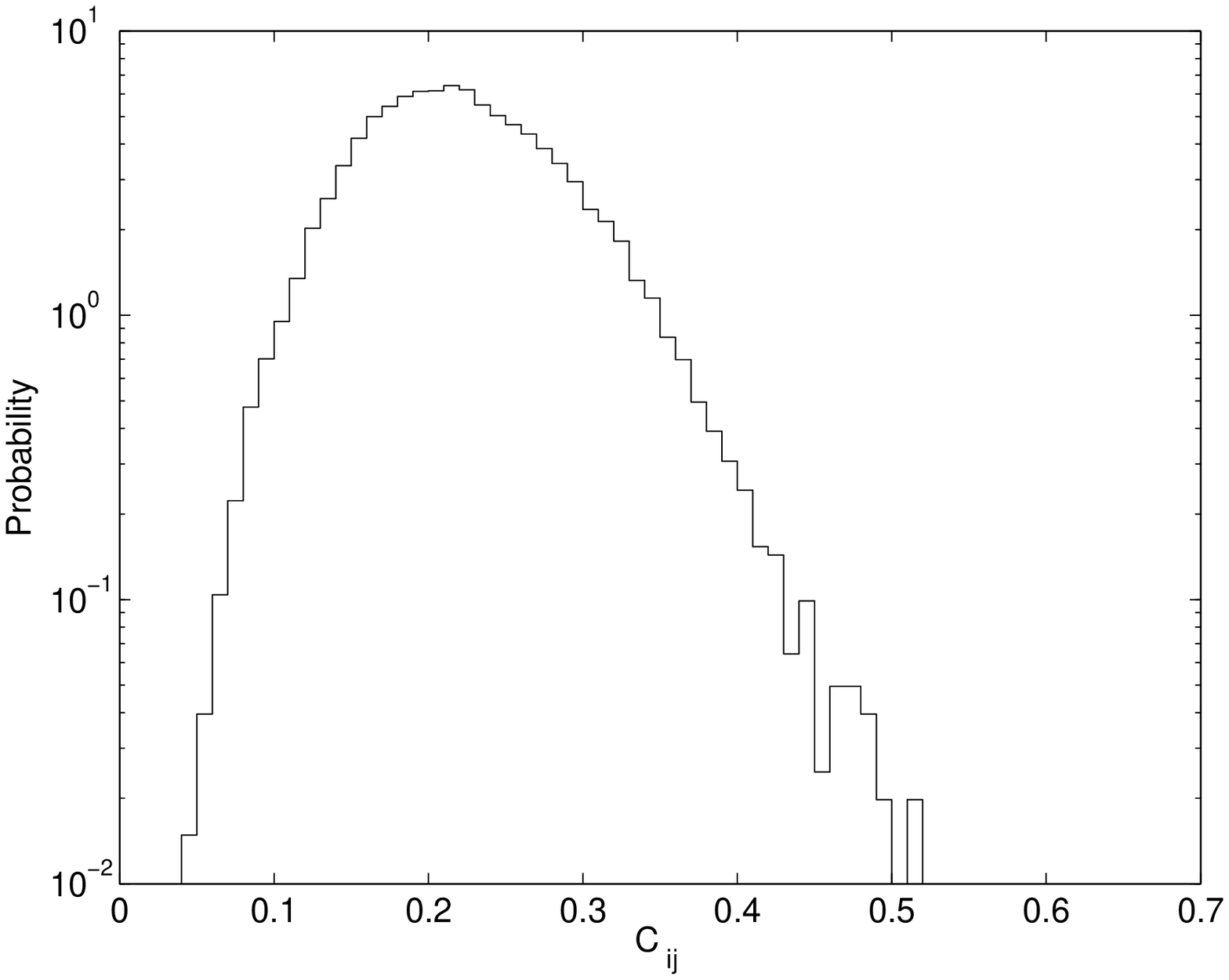}
\includegraphics[width=0.49\linewidth,clip]{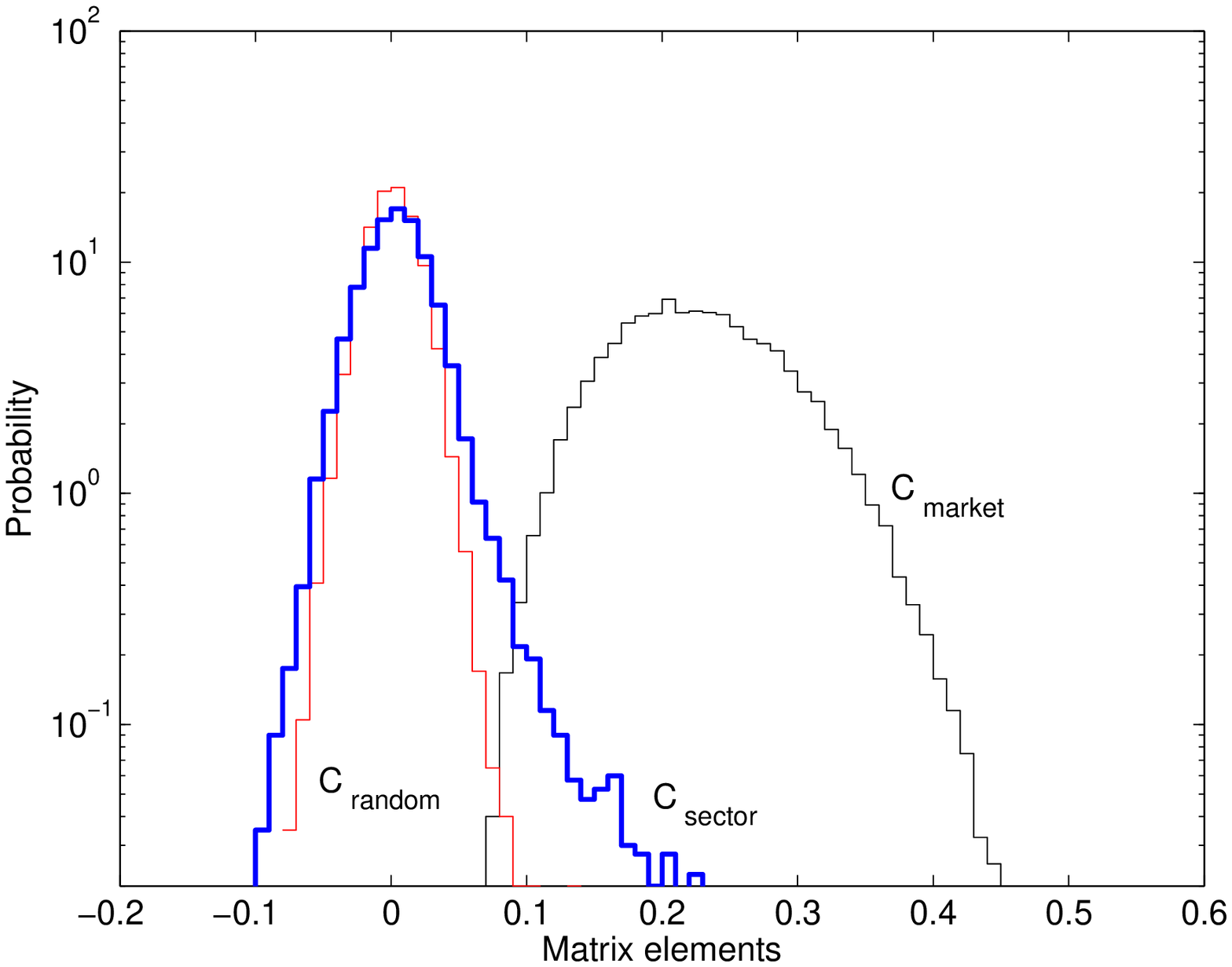}\\
\caption{(left) The distribution for the components $C_{ij}$ of the
cross-correlation matrix for NSE. (Right) The matrix element distributions
following decomposition of ${\mathbf C}$ into sector, $\mathbf {C}_{sector}$, 
market, ${\mathbf C}_{market}$, and random effects, ${\mathbf C}_{random}$, 
with $N_s=5$.}
\label{ss:compare}       
\end{figure}

Therefore, to be able to identify the internal structure of interactions
between the stocks we need to remove the market mode, i.e., the
effect of the largest eigenvalue. Also, the effect of random noise has to
be filtered out.
To perform this filtering, we use the method proposed in Ref.~\cite{kim05}
where the correlation matrix was expanded in terms of its eigenvalues
$\lambda_{i}$ and the corresponding eigenvectors ${\mathbf u}_{i}$:
${\mathbf C}= \Sigma_i \lambda_i \mathbf{u}_i \mathbf{u}_i^T$. This allows the correlation matrix
to be decomposed into three parts, corresponding to the market, sector
and random components:
\begin{equation}
{\mathbf C} = {\mathbf C}_{market} + {\mathbf C}_{sector} + 
{\mathbf C}_{random} = \lambda_0 \mathbf{u}_0^T \mathbf{u}_0 + \sum_{i =1}^{N_{s}} 
 \mathbf{u}_i^T \mathbf{u}_i + \sum_{i = N_{s}+1}^{N-1} \mathbf{u}_i^T
 \mathbf{u}_i,
\end{equation}
where, the eigenvalues have been arranged in descending order (the largest
labelled 0) and $N_{s}$ is the number of intermediate eigenvalues.
From the empirical data, it is not often obvious what is the value of 
$N_{s}$, as the bulk may deviate from the predictions of random matrix
theory because of underlying structure induced correlations. For this
reason, we use visual inspection of the distribution to choose $N_{s}
= 5$, and verify that small changes in this value does not alter the results.
The robustness of our results to small variations in the estimation of
$N_{s}$ is because the error involved is only due to the
eigenvalues closest to the bulk that have the smallest contribution to 
${\mathbf C}_{sector}$. Fig.~\ref{ss:compare}
shows the result of the decomposition
of the full correlation matrix into the three components.
Compared to the NYSE, NSE shows
a less extended tail for the sector correlation matrix elements 
$C_{ij}^{sector}$. This implies that the Indian market has a much smaller
fraction of strongly interacting stocks, which would be the case if there
is no significant segregation into sectors in the market.

\begin{figure}[tbp] \centering
\includegraphics[width=0.8\linewidth,clip]{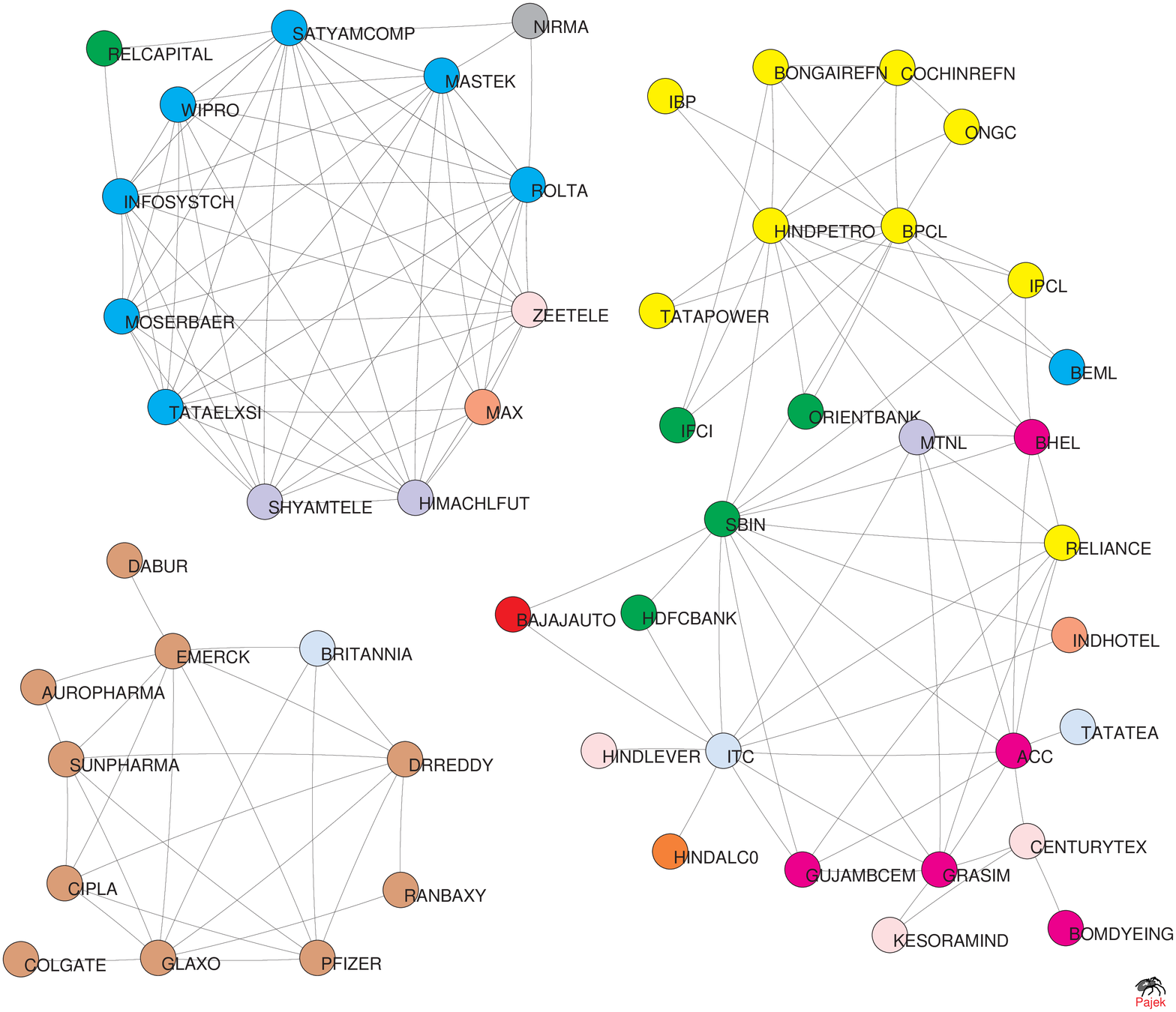}
\caption{The network of stock interactions in NSE
generated from the group correlation matrix ${\mathbf C}_{sector}$ 
with threshold $c^{*} = 0.09$. The node colors indicate the business
sector to which a stock belongs. The top
left cluster comprises mostly Technology stocks, while the bottom left
cluster is composed almost entirely of 
Healthcare \& Pharmaceutical stocks. By contrast, the larger cluster on
the right is not dominated by any particular sector.
The figure has been drawn using the Pajek software.}
\label{ss:network}       
\end{figure}
Next, we construct the network of interactions among stocks by using the
information in the sector correlation matrix~\cite{kim05}. The 
binary-valued adjacency 
matrix ${\mathbf A}$ of the network is generated from 
${\mathbf C}_{sector}$ by using a threshold $c_{th}$ such that $A_{ij} = 1$
if $C_{ij}^{sector} > c_{th}$, $A_{ij} = 0$ otherwise. If the long tail in the
$C_{ij}^{sector}$ distribution is indeed due to correlations among stocks
belonging to a particular business sector, this should be reflected in
a clustered structure of the network for an appropriate choice of the threshold.
Fig.~\ref{ss:network} shows the resultant network for the best choice of
$c_{th}=c^*$ (= 0.09) in terms of creating the largest clusters of related stocks. 
However, even for the ``best'' choice we find that only two sectors have
been properly clustered, those corresponding to Technology and to Pharmaceutical
Companies. The majority of the frequently traded stocks cannot be arranged
into well-segregated groups corresponding to the various business sectors
they belong to. This failure again reflects the fact that intra-group
correlations in most cases are much weaker compared to the market-wide
correlation in the Indian market.


\section{Time-evolution of the Correlation Structure}
\label{sec:5}
\begin{figure}[tbp] \centering
\includegraphics[width=0.9\linewidth,clip]{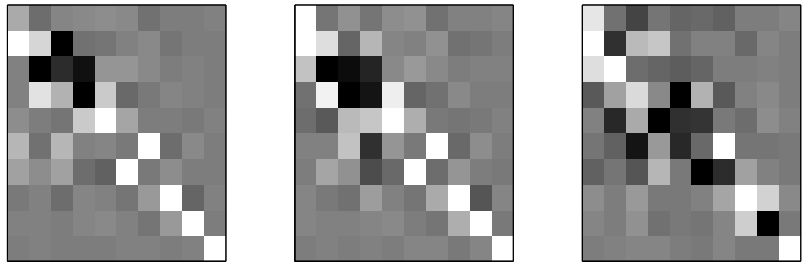}
\caption{Grayscale pixel representation of the overlap matrix 
as a function of
time for daily data during the period 1996-2001 taken as the
reference. Here, the gray scale
coding is such that white corresponds to $O_{ij} = 1$ and black
corresponds to $O_{ij} = 0$. The length of the time window used to
compute C is $T = 1250$ days (5 years) and the separations used to calculate
$O_{ij}$ are $\tau = 6$ months (left), 1 year (middle) and 2 years (right).
The diagonal represents the overlap between the components of
the corresponding eigenvectors for the 10 largest eigenvalues of the
original and shifted windows. The bottom right corner corresponds to
the largest eigenvalue.}
\label{ss:overlap}       
\end{figure}
In this section, we study the temporal properties of the correlation matrix.
We note here that if the deviations from the random matrix
predictions are indicators of genuine correlations, then the 
eigenvectors corresponding to the deviating eigenvalues should be
stable in time,
over the period used to calculate the correlation matrix.
We choose the eigenvectors corresponding to the 10 largest eigenvalues
for the correlation matrix over a period $A = [t, t+T]$  to construct 
a $10 \times 201$ matrix ${\mathbf D}_A$. A similar matrix ${\mathbf D}_B$ can 
be generated by using a different time period $B = [t + \tau, t + \tau + T]$
having the same duration but a time lag $\tau$ compared to the other. 
These are then 
used to generate the $10 \times 10$ overlap matrix ${\mathbf O}(t, \tau)$ = 
$\mathbf{D}_A \mathbf{D}_B^T$. In the ideal case, when the 10 eigenvectors are
absolutely stable in time, ${\mathbf O}$ would be a identity matrix.
For the NSE data we have used time lags of $\tau$ = 6 months, 1 year and 
2 years, for a time window of 5 years and the reference period beginning
in Jan 1996. As shown in Fig.~\ref{ss:overlap} the eigenvectors show
different degrees of stability, with the one corresponding to the largest
eigenvalue being the most stable. The remaining eigenvectors show
decreasing stability with an increase in the lag period.

\begin{figure}[tbp] \centering
\includegraphics[width=0.9\linewidth,clip]{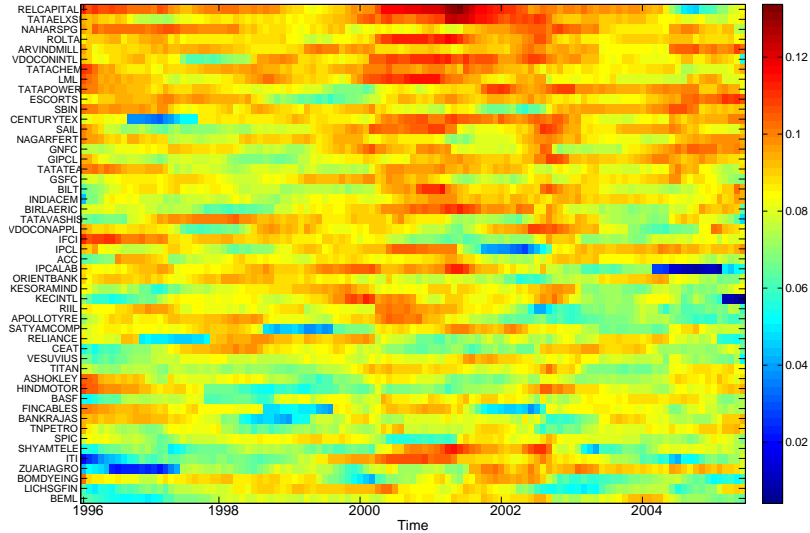}
\caption{The 50 stocks which have the largest contribution to the
eigenvector components of the largest eigenvalue
as a function of time for the period Jan 1996-May 2006.
The color
intensity represents the degree of correlation.}
\label{ss:startend}       
\end{figure}
Next, we focus on the temporal evolution of the composition of the eigenvector
corresponding to the {\em largest} eigenvalue. Our purpose is to find the set
of stocks that have consistently high contributions to this eigenvector,
and they can be identified as the ones whose behavior is dominating
the market mode. We study the time-development
by dividing the return time-series data into $M$ 
overlapping sets of length $T$. Two
consecutive sets are displaced relative to each other by a time lag 
$\delta t$. In our study,
$T$ is taken as six months (125 trading days), while $\delta t$ is taken
to be one
month (21 trading days). 
The resulting correlation matrices, ${\mathbf C}_{T,\delta t}$, can now be 
analysed to get further
understanding of the time-evolution of correlated movements among 
the different stocks.

In a previous paper~\cite{sinha06}, 
we have found that the largest eigenvalue of 
${\mathbf C}_{T,\delta t}$  
follows closely the time variation of the average correlation coefficient.
This indicates that the largest eigenvalue $\lambda_{0}$
captures the behavior
of the entire market. However, the relative contribution to its eigenvector
$\mathbf{u}_{0}$
by the different stocks may change over time. We assume that if a company
is a really important player in the market, then it will have a significant
contribution in the composition of $\mathbf{u}_{0}$ over many time windows.
Fig.~\ref{ss:startend} shows the 50 largest stocks in terms of
consistently having large representation in $\mathbf{u}_{0}$.
Note the existence of 5 companies from the Tata group and 3 companies
of the Reliance group in this set. This is consistent with the
general belief in the business community that these two groups dominate
the Indian market, and may disproportionately affect the market through
their actions.

\section{Conclusions}
In this paper, we have examined the structure of the Indian financial
market through a detailed investigation of the spectral properties of the
cross-correlation matrix of price returns.
We demonstrate that the eigenvalue distribution
is similar to that observed for developed markets of USA and Japan. However,
unlike the latter, the Indian market shows much less evidence of the existence
of business sectors having distinct identities. In fact, most of the
observed correlation among stocks is due to effects common to the entire
market, which has the effect of making the Indian market appear more
correlated than developed markets. We hypothesise that the reason why
emerging markets have been often reported to be significantly more correlated
is because they are distinguished from developed ones in the absence of
strong interactions between clusters of stocks in the former.
This has implications for the understanding of markets as complex interacting
systems, namely, that interactions emerge between groups of stocks
as a market evolves over time to finally exhibit the clustered structure
characterizing, e.g., the NYSE. How such self-organization is related to
other changes a market undergoes as it develops is a question worth 
pursuing with the tools available to econophysicists. From the point of
view of possible applicability, these results are of significance to
the problem of portfolio diversification. With the advent of liberalization,
there has been a significant flow of investment into the Indian market.
The question of how investments can be made over a balanced portfolio of stocks
so as to minimize risks assumes importance in such a situation. Our study
indicates that schemes for constructing such optimized portfolios must take
into account the fact that emerging markets are in general less differentiated
and more correlated than developed markets.

\begin{table}
\centering
\caption{The list of 201 stocks in NSE analyzed in this paper.}

\begin{scriptsize}
\begin{tabular}{cll |@{\mbox{ }} cll}\hline\hline
 $i$ & Company & Sector & $i$ & Company & Sector \\ \hline
   1 & UCALFUEL   & Automobiles Transport &  61 & SUPPETRO   & Energy           \\  
   2 & MICO	  & Automobiles Transport &  62 & DCW	     & Energy           \\  
   3 & SHANTIGEAR & Automobiles Transport &  63 & CHEMPLAST  & Energy           \\  
   4 & LUMAXIND   & Automobiles Transport &  64 & RELIANCE   & Energy           \\  
   5 & BAJAJAUTO  & Automobiles Transport &  65 & HINDPETRO  & Energy           \\  
   6 & HEROHONDA  & Automobiles Transport &  66 & BONGAIREFN & Energy           \\  
   7 & MAHSCOOTER & Automobiles Transport &  67 & BPCL	     & Energy           \\  
   8 & ESCORTS    & Automobiles Transport &  68 & IBP	     & Energy           \\  
   9 & ASHOKLEY   & Automobiles Transport &  69 & ESSAROIL   & Energy           \\  
  10 & M\&M	  & Automobiles Transport &  70 & VESUVIUS   & Energy           \\  
  11 & EICHERMOT  & Automobiles Transport &  71 & NOCIL	     & Basic Materials  \\
  12 & HINDMOTOR  & Automobiles Transport &  72 & GOODLASNER & Basic Materials  \\
  13 & PUNJABTRAC & Automobiles Transport &  73 & SPIC	     & Basic Materials  \\
  14 & SWARAJMAZD & Automobiles Transport &  74 & TIRUMALCHM & Basic Materials  \\
  15 & SWARAJENG  & Automobiles Transport &  75 & TATACHEM   & Basic Materials  \\
  16 & LML        & Automobiles Transport &  76 & GHCL	     & Basic Materials  \\
  17 & VARUNSHIP  & Automobiles Transport &  77 & GUJALKALI  & Basic Materials  \\
  18 & APOLLOTYRE & Automobiles Transport &  78 & PIDILITIND & Basic Materials  \\
  19 & CEAT       & Automobiles Transport &  79 & FOSECOIND  & Basic Materials  \\
  20 & GOETZEIND  & Automobiles Transport &  80 & BASF	     & Basic Materials  \\
  21 & MRF	  & Automobiles Transport &  81 & NIPPONDENR & Basic Materials  \\
  22 & IDBI	  & Financial             &  82 & LLOYDSTEEL & Basic Materials  \\
  23 & HDFCBANK   & Financial             &  83 & HINDALC0   & Basic Materials  \\
  24 & SBIN       & Financial             &  84 & SAIL	     & Basic Materials  \\
  25 & ORIENTBANK & Financial             &  85 & TATAMETALI & Basic Materials  \\
  26 & KARURVYSYA & Financial             &  86 & MAHSEAMLES & Basic Materials  \\
  27 & LAKSHVILAS & Financial             &  87 & SURYAROSNI & Basic Materials  \\
  28 & IFCI       & Financial             &  88 & BILT	     & Basic Materials  \\
  29 & BANKRAJAS  & Financial             &  89 & TNPL	     & Basic Materials  \\
  30 & RELCAPITAL & Financial             &  90 & ITC	     & Consumer Goods   \\
  31 & CHOLAINV   & Financial             &  91 & VSTIND     & Consumer Goods   \\
  32 & FIRSTLEASE & Financial             &  92 & GODFRYPHLP & Consumer Goods   \\
  33 & BAJAUTOFIN & Financial             &  93 & TATATEA    & Consumer Goods   \\
  34 & SUNDARMFIN & Financial             &  94 & HARRMALAYA & Consumer Goods   \\
  35 & HDFC       & Financial             &  95 & BALRAMCHIN & Consumer Goods   \\
  36 & LICHSGFIN  & Financial             &  96 & RAJSREESUG & Consumer Goods   \\
  37 & CANFINHOME & Financial             &  97 & KAKATCEM   & Consumer Goods   \\
  38 & GICHSGFIN  & Financial             &  98 & SAKHTISUG  & Consumer Goods   \\
  39 & TFCILTD    & Financial             &  99 & DHAMPURSUG & Consumer Goods   \\
  40 & TATAELXSI  & Technology            & 100 & BRITANNIA  & Consumer Goods   \\
  41 & MOSERBAER  & Technology            & 101 & SATNAMOVER & Consumer Goods       \\
  42 & SATYAMCOMP & Technology            & 102 & INDSHAVING & Consumer Goods       \\
  43 & ROLTA      & Technology            & 103 & MIRCELECTR & Consumer Discretonary\\
  44 & INFOSYSTCH & Technology            & 104 & SURAJDIAMN & Consumer Discretonary\\
  45 & MASTEK     & Technology            & 105 & SAMTEL     & Consumer Discretonary\\
  46 & WIPRO      & Technology            & 106 & VDOCONAPPL & Consumer Discretonary\\
  47 & BEML       & Technology            & 107 & VDOCONINTL & Consumer Discretonary\\
  48 & ALFALAVAL  & Technology            & 108 & INGERRAND  & Consumer Discretonary\\
  49 & RIIL       & Technology            & 109 & ELGIEQUIP  & Consumer Discretonary\\
  50 & GIPCL      & Energy                & 110 & KSBPUMPS   & Consumer Discretonary\\
  51 & CESC       & Energy                & 111 & NIRMA	     & Consumer Discretonary\\ 
  52 & TATAPOWER  & Energy                & 112 & VOLTAS     & Consumer Discretonary\\ 
  53 & GUJRATGAS  & Energy                & 113 & KECINTL    & Consumer Discretonary\\ 
  54 & GUJFLUORO  & Energy                & 114 & TUBEINVEST & Consumer Discretonary\\ 
  55 & HINDOILEXP & Energy                & 115 & TITAN	     & Consumer Discretonary\\ 
  56 & ONGC	  & Energy                & 116 & ABB	     & Industrial           \\ 
  57 & COCHINREFN & Energy                & 117 & BHEL	     & Industrial           \\ 
  58 & IPCL	  & Energy                & 118 & THERMAX    & Industrial           \\ 
  59 & FINPIPE    & Energy                & 119 & SIEMENS    & Industrial           \\ 
  60 & TNPETRO    & Energy                & 120 & CROMPGREAV & Industrial           \\ 
\hline \hline
\end{tabular}      
\end{scriptsize}
\label{ss:table1}
\end{table}

\begin{scriptsize}
\begin{tabular}{cll |@{\mbox{ }} cll}\hline\hline
 $i$ & Company & Sector & $i$ & Company & Sector \\ \hline
 121 & HEG	  & Industrial            & 161 & HIMACHLFUT & Telecom       \\ 
 122 & ESABINDIA  & Industrial            & 162 & MTNL	     & Telecom       \\ 
 123 & BATAINDIA  & Industrial            & 163 & BIRLAERIC  & Telecom       \\ 
 124 & ASIANPAINT & Industrial            & 164 & INDHOTEL   & Services      \\ 
 125 & ICI	  & Industrial            & 165 & EIHOTEL    & Services      \\ 
 126 & BERGEPAINT & Industrial            & 166 & ASIANHOTEL & Services      \\ 
 127 & GNFC	  & Industrial            & 167 & HOTELEELA  & Services      \\ 
 128 & NAGARFERT  & Industrial            & 168 & FLEX	     & Services      \\ 
 129 & DEEPAKFERT & Industrial            & 169 & ESSELPACK  & Services      \\ 
 130 & GSFC	  & Industrial            & 170 & MAX	     & Services      \\ 
 131 & ZUARIAGRO  & Industrial            & 171 & COSMOFILMS & Services      \\
 132 & GODAVRFERT & Industrial            & 172 & DABUR	     & Health Care   \\
 133 & ARVINDMILL & Industrial            & 173 & COLGATE    & Health Care   \\
 134 & RAYMOND    & Industrial            & 174 & GLAXO	     & Health Care   \\
 135 & HIMATSEIDE & Industrial            & 175 & DRREDDY    & Health Care   \\
 136 & BOMDYEING  & Industrial            & 176 & CIPLA	     & Health Care   \\
 137 & NAHAREXP   & Industrial            & 177 & RANBAXY    & Health Care   \\
 138 & MAHAVIRSPG & Industrial            & 178 & SUNPHARMA  & Health Care   \\
 139 & MARALOVER  & Industrial            & 179 & IPCALAB    & Health Care   \\
 140 & GARDENSILK & Industrial            & 180 & PFIZER     & Health Care   \\
 141 & NAHARSPG   & Industrial            & 181 & EMERCK     & Health Care   \\
 142 & SRF	  & Industrial            & 182 & NICOLASPIR & Health Care   \\
 143 & CENTENKA   & Industrial            & 183 & SHASUNCHEM & Health Care   \\
 144 & GUJAMBCEM  & Industrial            & 184 & AUROPHARMA & Health Care   \\
 145 & GRASIM     & Industrial            & 185 & NATCOPHARM & Health Care   \\
 146 & ACC	  & Industrial            & 186 & HINDLEVER  & Miscellaneous \\
 147 & INDIACEM   & Industrial            & 187 & CENTURYTEX & Miscellaneous \\
 148 & MADRASCEM  & Industrial            & 188 & EIDPARRY   & Miscellaneous \\
 149 & UNITECH    & Industrial            & 189 & KESORAMIND & Miscellaneous \\
 150 & HINDSANIT  & Industrial            & 190 & ADANIEXPO  & Miscellaneous \\
 151 & MYSORECEM  & Industrial            & 191 & ZEETELE    & Miscellaneous \\
 152 & HINDCONS   & Industrial            & 192 & FINCABLES  & Miscellaneous \\
 153 & CARBORUNIV & Industrial            & 193 & RAMANEWSPR & Miscellaneous \\
 154 & SUPREMEIND & Industrial            & 194 & APOLLOHOSP & Miscellaneous \\
 155 & RUCHISOYA  & Industrial            & 195 & THOMASCOOK & Miscellaneous \\
 156 & BHARATFORG & Industrial            & 196 & POLYPLEX   & Miscellaneous \\
 157 & GESHIPPING & Industrial            & 197 & BLUEDART   & Miscellaneous \\
 158 & SUNDRMFAST & Industrial            & 198 & GTCIND     & Miscellaneous \\
 159 & SHYAMTELE  & Telecom               & 199 & TATAVASHIS & Miscellaneous \\
 160 & ITI	  & Telecom               & 200 & CRISIL     & Miscellaneous \\
     &            &                       & 201 & INDRAYON   & Miscellaneous \\
\hline \hline
\end{tabular}      
\end{scriptsize}

\begin{table}
\centering
\caption{Stocks with dominant contribution to the six smallest eigenvalues.}

\begin{scriptsize}

\begin{tabular}{cll @{\mbox{ }} cll}\hline\hline
 $\lambda_{201}$ & $\lambda_{200}$ & $\lambda_{199}$ & $\lambda_{198}$ &
 $\lambda_{197}$ & $\lambda_{196}$ \\ \hline
SBIN      & SBIN       & RELCAPITAL & RELCAPITAL &  HINDPETRO  & HINDPETRO  \\ 
TATAELXSI & ORIENTBANK & VDOCONAPPL & BPCL       &  BPCL       & BPCL       \\ 
ROLTA     & TATAELXSI  & VDOCONINTL & VDOCONAPPL &  VDOCONINTL & GNFC       \\ 
          & ROLTA      &            & VDOCONINTL &  GNFC       & GSFC       \\ 
          & ACC        &            & NAHARSPG   &  NAHARSPG   & NAHAREXP   \\ 
          &            &            &            &             &  NAHARSPG  \\ 
          &            &            &            &             &  ESSELPACK \\
\hline \hline
\end{tabular}      
\end{scriptsize}
\label{ss:table2}
\end{table}

\vspace{0.1cm}
\noindent
{\small {\bf Acknowledgements:} We thank N.~Vishwanathan for
assistance in preparing the data for analysis and M.~Marsili for 
helpful discussions.}


\printindex
\end{document}